# One-dimensional van der Waals heterostructures

Rong Xiang[1,12,*], Taiki Inoue[1,12], Yongjia Zheng[1,12] Akihito Kumamoto[2], Yang Qian[1], Yuta Sato[3], Ming Liu[1], Devashish Gokhale[4], Jia Guo[1,5], Kaoru Hisama[1], Satoshi Yotsumoto[1], Tatsuro Ogamoto[1], Hayato Arai[1], Yu Kobayashi[6], Hao Zhang[1], Bo Hou[7], Anton Anisimov[8], Yasumitsu Miyata[6], Susumu Okada[9], Shohei Chiashi[1], Yan Li[1,5], Jing Kong[10], Esko I. Kauppinen[11], Yuichi Ikuhara[2], Kazu Suenaga[3], Shigeo Maruyama[1,7,*]

[1] *Departure of Mechanical Engineering, The University of Tokyo, Tokyo 113-8656, Japan*

[2] *Institute of Engineering Innovation, The University of Tokyo, Tokyo 113-8656, Japan*

[3] *Nanomaterials Research Institute, National Institute of Advanced Industrial Science and Technology (AIST), Tsukuba 305-8565, Japan*

[4] *Department of Chemical Engineering, Indian Institute of Technology Madras, Chennai 600036, India*

[5] *College of Chemistry and Molecular Engineering, Peking University, Beijing 100871, China*

[6] *Department of Physics, Tokyo Metropolitan University, Tokyo 192-0397, Japan*

[7] *Energy Nano Engineering Lab., National Institute of Advanced Industrial Science and Technology (AIST), Tsukuba 305-8564, Japan*

[8]*Canatu Ltd., Helsinki FI‐00390, Finland*

[9] *Graduate School of Pure and Applied Sciences, University of Tsukuba, Tsukuba 305-8571, Japan*

[10] *Department of Electrical Engineering and Computer Science, Massachusetts Institute of Technology, Massachusetts 02139, United States*

[11] *Department of Applied Physics, Aalto University School of Science, Espoo 15100, FI-00076 Aalto, Finland*

[12] *These authors contributed equally. Rong Xiang, Taiki Inoue, Yongjia Zheng*

*email: xiangrong@photon.t.u-tokyo.ac.jp; maruyama@photon.t.u-tokyo.ac.jp

Property by design is one appealing idea in material synthesis but hard to achieve in practice. A recent successful example is the demonstration of van der Waals (vdW) heterostructures,[1-3] in which atomic layers are stacked on each other and different ingredients can be combined beyond symmetry and lattice matching. This concept, usually described as a nanoscale Lego blocks, allows to build sophisticated structures layer by layer. However, this concept has been so far limited in two dimensional (2D) materials. Here we show a class of new material where different layers are coaxially (instead of planarly) stacked. As the structure is in one dimensional (1D) form, we name it "1D vdW heterostructures". We demonstrate a 5 nm diameter nanotube consisting of three different materials: an inner conductive carbon nanotube (CNT), a middle insulating hexagonal boron nitride nanotube (BNNT) and an outside semiconducting $MoS_2$ nanotube. As the technique is highly applicable to other materials in the current 2D libraries,[4-6] we anticipate our strategy to be a starting point for discovering a class of new semiconducting nanotube materials. A plethora of function-designable 1D heterostructures will appear after the combination of CNTs, BNNTs and semiconducting nanotubes.

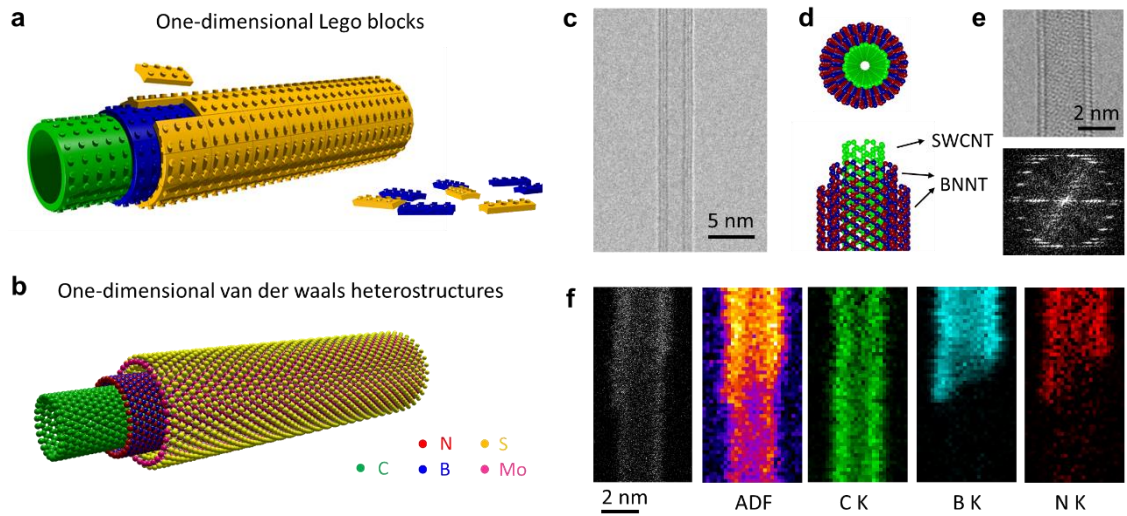

**Figure 1 Overview of 1D vdW heterostructures.** (a) A one dimensional Lego block showing the concept of this work: atomic layers of different materials can be coaxially stacked to form a 1D vdW heterostructure; (b) the atomic model of one material built in this study: metal-insulator-semiconductor (M+I+S) 1D vdW heterostructures; (c) TEM image and (d) atomic models of a SWCNT wrapped with two layers of BNNT; (e) aberration-corrected TEM image of a SWCNT-BNNT and its fast Fourier transform (FFT); (f) EELS mapping of a SWCNT partially wrapped with BN nanotube showing the inner is carbon and outer is BN.

When playing Lego blocks, one usually starts from a green baseplate, on which various structures are built. In this study, the baseplate is single-walled carbon nanotube (SWCNT).[7] SWCNT is chosen as this starting material for several reasons. First and foremost it is so far the best-studied 1D material and can be synthesized in many controlled geometries. Second, SWCNT can be a good conductor so it may serve as the electrode for a future device. The SWCNT film (which we used in most of the following demonstrations) has a sheet resistance about 80 ohm/sq. at 90%, comparable to the indium tin oxide (ITO) films that are widely used in modern technology.[8] The schematic of 1D Lego blocks and vdW heterostructures are presented in Figure 1a-b.

First, we present a SWCNT-BNNT 1D heterostructure before $MoS_2$ and other structures. We utilized SWCNTs as a template and synthesized additional hexagonal BN layers by chemical vapor deposition (CVD). Figure 1c shows a representative high resolution transmission electron microscope (HRTEM) image of this coaxial heterostructure and more images are provided as Figure S1a. From a conventional HRTEM image, this nanotube looks not distinguishable from a triple-walled pure carbon nanotube. The aberration-corrected HRTEM image of a similar tube reveals a contrast of stacking of two perfect nanotubes (Figure 1d, e). However, as the starting material is purely single-walled before we performed a post BN coating, it is reasonable to expect that the outer wall(s) are BN. This point is evidenced by electron energy loss spectroscopic (EELS) mapping shown in Figure 1f. As the reaction occurs on the outer surface, different from previous attempts inside a nanotube,[9,10] continuous coating and perfectly crystalized outer BNNTs are achieved in this work. The number of outer BNNT walls can be adjusted from 1-2 to 5-8, depending on the time of BN CVD (Figure S1b,c).

This coaxial heterostructure can be characterized by many other techniques. X-ray photoelectron spectroscopy (XPS) reveals B-N bond and C-C bond in this sample but no apparent peaks of C-N and B-C, suggesting there is no noticeable substitution of atoms (Figure S2); optical absorption spectra (Figure S3a) reveals an emerge of peak near 205 nm and the peak intensity increases with extending growth time (number of outside BNNT layers); cathode luminance spectra (Figure S3b) of the sample confirm the existence of outer BN by emissions at UV range; Raman spectra of the SWCNT-BN (Figure S3c,d) show clear features for the inner SWCNTs, evidencing the high quality is preserved (this will be further supported later). Also, the growth strategy applies to SWCNTs with different morphologies, including vertical arrays, horizontal arrays, random networks and suspended SWCNTs bridging pillars (Figure S4).

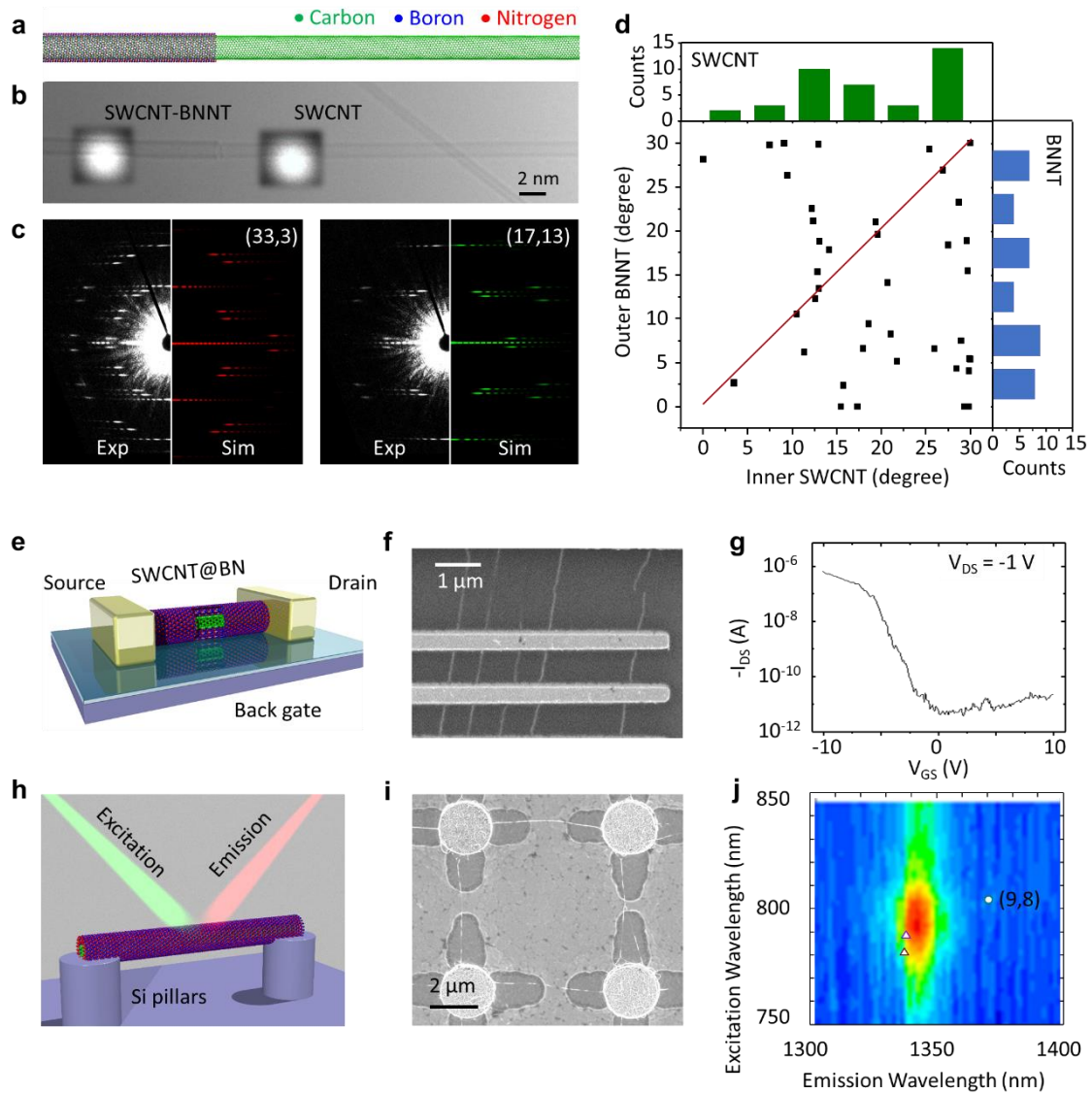

**Figure 2 Structure and property of SWCNT-BNNT vdW heterostructures.** (a) Atomic model and (b) TEM image of SWCNT-BNNT steps; (c) experimental and simulated ED pattern of the inner (16,14) SWCNT and outer (33,3) BNNT; (d) Plot of chiral angle of inner SWCNT vs. the outer BNNT for double-walled SWCNT-BNNT, revealing as-grown SWCNTs are enriched in armchair side but the outer BNNT is evenly distributed; (e) a schematic (f) SEM image, and (g) characteristic transfer curve of a FET built on a SWCNT-BNNT arrays; (g) a schematic and (i) SEM image of SWCNT-BN nanotube grown bridging the pillars; (j) PL excitation-emission map of as suspended (9,8) SWCNT after BN CVD.

The formation of this SWCNT-BNNT heterostructure follows an open-end growth mechanism. (Movie S1) As the outside BNNT is post-grown, the extension of

BNNT happens only at the open edge of BNNT. This is similar to the growth of an additional layer in 2D material, but very rarely observed in previous growth of carbon nanotubes.[11] This open edge growth mechanism is supported by the many growth step edges observed in an intermediated growth condition (Figure 2a-c). These steps also allow us to take nano-area electron diffraction (NED) patterns[12] of the pristine SWCNT and the same tube after BNNT growth. In this case, a (17,13) SWCNT is wrapped by a (33, 3) single-walled BNNT (Another example shown in Figure S5). Even without the steps, the 2% of lattice difference between SWCNT and BNNT [13] allows to distinguish them in the ED patterns. We performed chiral angle analysis shown (Figure S6, S7, Table S1). In a collection of 74 SWCNTs and 40 SWCNT-BNNT double-walled nanotubes, SWCNTs are enriched in near armchair side. This armchair enrichment is consistent with previous experimental and theoretical analysis.[14,15] However, the outer BNNTs are evenly distributed (even slightly prefers zigzag side). This distinct difference may be attributed the open-end growth model, which is different from the catalytic growth of pristine SWCNT.[15] Nonetheless, it is clearly no chiral angle dependence in this SWCNT-BN heterostructures, evidencing symmetry and lattice matching are not the limitation for material combination.[16]

Perfect BNNT coating does not change the intrinsic transport property of inner SWCNT. A field effect transistor (FET) built on this BNNT wrapped SWCNT show a similarly high performances with the ON/OFF ratio of $10^5$ (Figure 2e-g). Due to the large band-gap of BNNT, photoluminescence (PL) of inner SWCNT is also preserved after BN CVD (Figure 2h-j). [17,18] Additionally, BN wrapping increases the thermal stability of SWCNT. Pure SWCNTs are well known to burn in air at a temperature around 450°C. After coating, the structure survives at up to 700°C (Figure S8). Various device and other functions may be expected for this perfectly crystalized heterostructures.

The successful post-stacking of BNNT onto the outer surface of SWCNTs makes us believe a similar strategy can be employed to build other more sophisticated structures, or even those materials never-existed. Single-walled $MoS_2$ nanotube is one example. $MoS_2$ 2D sheet is intensively studied as the representative of transmission metal dichalcogenide (TMD) material in recent years.[19] Single-walled $MoS_2$ nanotube is, however, not convincingly demonstrated in previous experiments. Figure 3a-c show the atomic structure, TEM, scanning TEM (STEM) images of SWCNT-$MoS_2$ coaxial nanotube obtained after applying the strategy. The $MoS_2$ nanotube distinguishes itself by a much stronger contrast than carbon in both TEM and STEM images. (More images and EELS mapping are shown Figure S9) This is as far as we know one of the first TEM images evidencing the existence of single-walled MoS$_2$ nanotube. We emphasize there single-walled MoS$_2$ nanotube because multi-walled tubes (usually 20 nm or above) and its hybrid materials have been known for decades.[20-22] Single-walled MoS2 nanotube is predicted to have direct band-gap so its properties can probably be distinctively different from its thicker multi-walled counterparts.[23] Also, quantum confinement is expected to be significant in a single-walled MoS$_2$ nanotube with this small diameter.[24] Most importantly, formation of a MoS$_2$ nanotube suggests that other single-walled TMD nanotubes, e.g. WS$_2$, WSe$_2$, may also be produced by a similar approach. Hence, for this single-layer feature, largely curved, quantum-confineable TMD nanotubes, we foresee lots of interesting research topics in spectroscopy, device, and others.

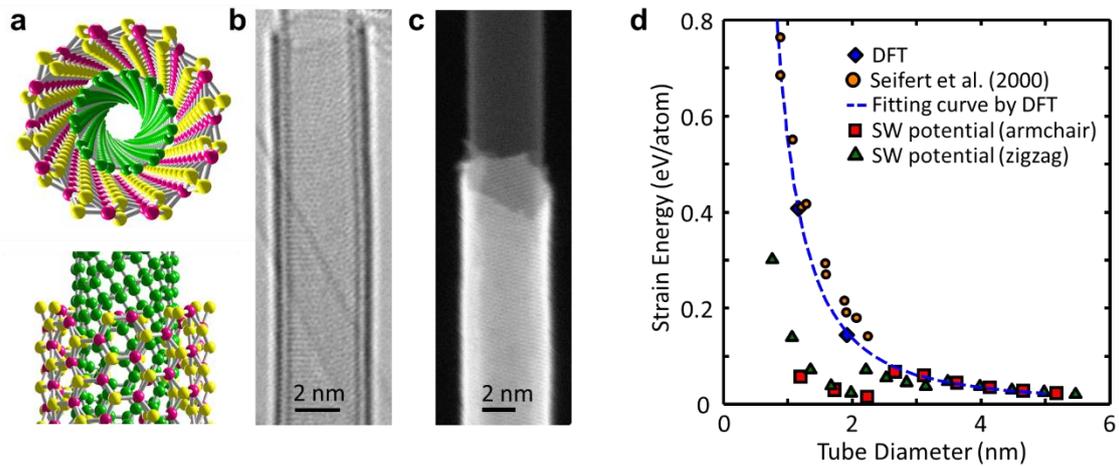

**Figure 3 SWCNT-MoS$_2$ 1D vdW heterostructure.** (a) Atomic model, (b) HRTEM image and (c) high angle annular dark field (HAADF) STEM image of a single-walled MoS$_2$ nanotube grown on a SWCNT; (d) the strain energy of a single-walled MoS$_2$ nanotube as a function of tube diameter calculated by SW potential and DFT simulation.

However, we observe a strong diameter dependence for the formation of single-walled MoS$_2$ nanotubes. Unlike BNNT wrapping, the yield of MoS$_2$ nanotube is very low and seamless wrapping is only observed on large diameter SWCNTs. To understand this point, we performed some simulations on the strain energy of single-walled MoS$_2$ nanotubes with different diameters. A clear $1/D^2$ relationship is observed in Figure 3d, which suggests that, in small diameter range, strain energy is significantly higher. This result is comparable to some previous attempts using density-functional-based tight binding (DFTB) available in literature.[23,25] If compared to SWCNTs,[26] this strain energy is much higher. This can be simply interpreted as a thickness effect: as single layer of MoS$_2$ contains three atomic planes and much unstable to roll into a tubular structure. This is probably why MoS$_2$ nanotubes were only seen as multi-walled or on multi-walled CNTs previously.[20,27] According to this result, the minimum diameter of single-walled MoS$_2$ nanotube should be significantly larger than SWCNTs. This prediction is surprisingly consistent with our experimental observations. The tubes (Figure S9c-f) have diameter ranging from 3.9 – 6.8 nm, only forming on SWCNTs that have diameter of 3 nm or more. However, most of the starting SWCNTs are thinner than 3 nm, which probably caused the low yield of SWCNT-MoS$_2$ heterostructures. In addition, there is no clear evidence showing preference of zigzag or armchair MoS$_2$ nanotube but we speculate their growth should be kinetically different. Besides this, many features of this single-walled MoS$_2$ nanotube are unknown, and we expect more efforts to come for this new material.

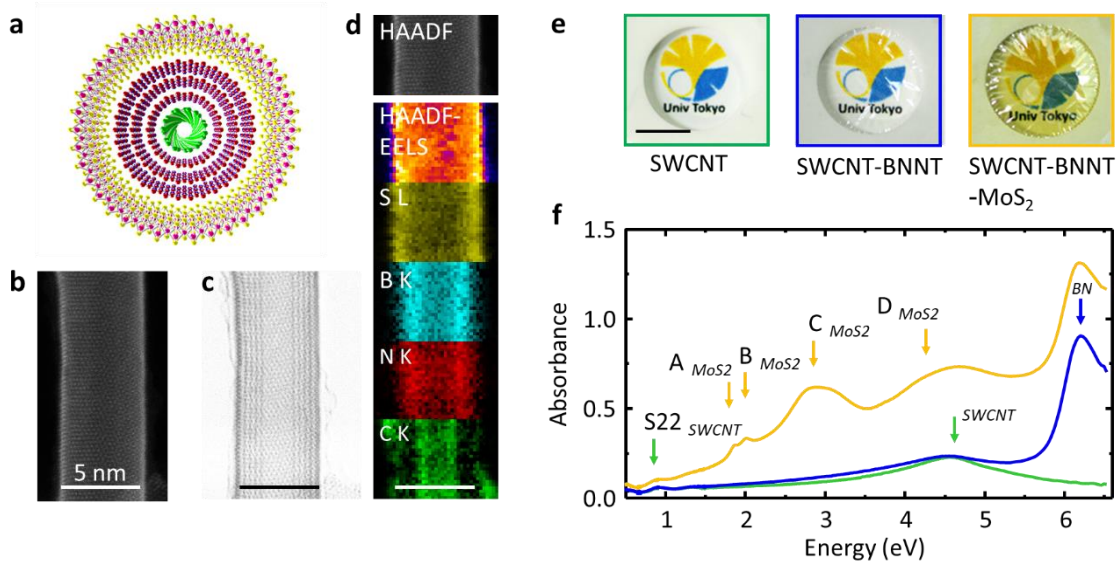

**Figure 4 SWNT-BNNT-MoS$_2$ 1D vdW heterostructures.** (a) atomic model, (b) HAADF-STEM image (c) annular bright field (ABF)-STEM image (scale bar 5 nm) and (d) EELS mapping of a 5 nm diameter ternary 1D vdW heterostructures consisting one layer of carbon, three layers of BNNT and one layer MoS$_2$ nanotube; (e) optical images of pristine SWCNT and coated nanotubes thin films (scale bar 5 mm) against a printed logo of the University of Tokyo; (f) corresponding optical absorption spectra of the different films, in which absorption from different materials can be distinguished.

Finally we demonstrate a ternary, SWCNT-BN-MoS$_2$ coaxial nanotube, shown in Figure 4a-d. It is 5 nm in diameter but consisting of three different materials: an inner one layer of conducting carbon nanotube, a middle three layers of insulating hexagonal BN and an outside one layer of semiconducting MoS$_2$. This structure is clearly visualized by the EELS mapping in Figure 4d. Also, as the tube diameter increases from 2 to more than 3 nm after BNNT coating, synthesizing an additional MoS$_2$ nanotube becomes much easier. More efficient growth are observed on most of the isolated regions. The coating can be produced in centimeter scale, and clear difference can be observed for original SWCNT, SWCNT-BNNT, SWCNT-BNNT-MoS$_2$ films even with naked eyes (Figure 4e). The optical absorption spectrum (Figure 4f) of the sample clearly reveals the photon absorption from three different layers. The EELS mapping of another nanotube is provided in Figure S9g. Such a structure can probably be the smallest metal-insulator-semiconductor device demonstrated so far. Field effect of the outside MoS$_2$ may be measured using the inner SWCNT gate, and photo-response, photovoltaic effect may also be expected in such 1D vdW heterostructures.

In conclusion, we have extended the concept of vdW heterostructures to a different dimension. It is highly possible that the synthetic technique shown here can be employed to generate a class of new nanotube materials and a series of more sophisticated combinations. One dimensional vdW heterostructures may bring numerous research interests in material synthesis, crystallography, optics, device physics, catalysis, and probably also some others that are still unforeseen.

**Acknowledgement**

Part of this work is financially supported by JSPS KAKENHI Grant Numbers JP25107002, JP15H05760, JP15K17984, JP16K05948, JP16H06333, 17K14601, 18H05329, IRENA Project by JST-EC DG RTD, Strategic International Collaborative Research Program, SICORP, and by the "Nanotechnology Platform" (project No. 12024046) of the Ministry of Education, Culture, Sports, Science and Technology (MEXT), Japan.

**Competing interests**

The authors declare no competing interests

**Author contributions:** R.X., T.I. and S.M. conceived and designed the experiments. E.K. and A.A. synthesized the SWCNT film. T.I., Y.Z., M.L. performed the BN CVD. Y.Q. performed the CVD growth of $MoS_2$ nanotubes using powder precursors. Y.M. and Y.K. performed the MOCVD growth of $MoS_2$ nanotubes. R.X. took the HRTEM images of heterostructures. A.K. took the EELS mapping of SWCNT-BN. Y.S. took the aberration corrected HRTEM image of SWCNT-BN, STEM image and EELS mapping of SWCNT-$MoS_2$. A.K. and R.X. took the STEM image and EELS mapping of SWCNT-BN-$MoS_2$ heterostructure. R.X. took the ED patterns of the SWCNT-BN nanotubes. R.X. and D.G. assigned the chirality of SWCNT-BN. D.G. analyzed the chiral angle distribution. A.K. simulated the ED pattern. T.I. fabricated the FET device and measured the transfer curve. H.A. synthesized the suspended SWCNTs. S.Y. and T.O measured the PL emission. J.G. studied the thermal stability of SWCNT-BN. T.I. measured and analyzed the XPS. Y.Z. and R.X. measured the CL. R.X. built the atomic structure of heterostructures. K.H. performed the SW potential and DFT simulation. R.X. wrote the paper. All authors joined the discussion and commented on the manuscript.

**Competing interests:** The authors declare no competing interests.

**Additional information**

Supplementary information is available for this paper at

**Correspondence and requests for materials** should be addressed to R.X. and S.M.